%% file: main-usenix.tex
\documentclass[letterpaper,twocolumn,10pt,table]{article}
\usepackage{usenix-2020-09}

\usepackage{xcolor}
\usepackage{framed}
\usepackage{multirow}
\usepackage{booktabs}
\usepackage{ifthen}
\usepackage{color}
\usepackage{amsmath}
\usepackage{xspace}
\usepackage[T1]{fontenc}
\usepackage{enumerate}
\usepackage[linesnumbered,ruled,vlined]{algorithm2e}
\usepackage{algorithm2e}
\usepackage{array,multirow,graphicx}
\usepackage{float}
\usepackage{balance}
\usepackage{tikz}
\usepackage{calc}
\usepackage{listings,amsfonts}
\usepackage{amsmath,pifont}
\usepackage{xspace}
\usepackage{hyperref,endnotes}
\usepackage{array}
\usepackage{multirow,makecell}
\usepackage[misc]{ifsym}
\usepackage{listings}
\usepackage{svg}
\usepackage[noend]{algpseudocode}
\usepackage{booktabs}
\usepackage[frozencache]{minted}
\usepackage{wrapfig}
\usepackage{caption}
\usepackage{subcaption}
\usepackage{mathtools}
\usepackage[most]{tcolorbox}

\usepackage{url}

\usepackage{breakurl}

\usepackage{longtable}

\setminted[java]{
    xleftmargin=15pt,
    framesep=2mm,
    linenos=true,
    breaklines=true,
    tabsize=2,
    encoding=utf8,
    frame=none,
    fontsize=\scriptsize,
    escapeinside=|| 
    }

\newcommand{\framework}{\textsc{AVVerifier}}
\newcommand{\grapher}{Grapher}
\newcommand{\simulator}{Simulator}
\newcommand{\detector}{Detector}

\begin{document}

\date{}

\title{All Your Tokens are Belong to Us: Demystifying Address Verification Vulnerabilities in Solidity Smart Contracts}

\author{
{\rm Tianle Sun}\\
Huazhong University of Science and Technology
\and
{\rm Ningyu He}\\
Peking University
\and
{\rm Jiang Xiao}\\
Huazhong University of Science and Technology
\and
{\rm Yinliang Yue}\\
Zhongguancun Laboratory
\and
{\rm Xiapu Luo}\\
The Hong Kong Polytechnic University
\and
{\rm Haoyu Wang*}\\
Huazhong University of Science and Technology
} 

\maketitle

\begin{abstract}
In Ethereum, the practice of verifying the validity of the passed addresses is a common practice, which is a crucial step to ensure the secure execution of smart contracts. Vulnerabilities in the process of address verification can lead to great security issues, and anecdotal evidence has been reported by our community. However, this type of vulnerability has not been well studied. To fill the void, in this paper, we aim to characterize and detect this kind of emerging vulnerability.
We design and implement {\framework}, a lightweight taint analyzer based on static EVM opcode simulation. Its three-phase detector can progressively rule out false positives and false negatives based on the intrinsic characteristics. Upon a well-established and unbiased benchmark, {\framework} can improve efficiency 2 to 5 times than the SOTA while maintaining a 94.3\% precision and 100\% recall. After a large-scale evaluation of over 5 million Ethereum smart contracts, we have identified 812 vulnerable smart contracts that were undisclosed by our community before this work, and 348 open source smart contracts were further verified, whose largest total value locked is over \$11.2 billion.
We further deploy {\framework} as a real-time detector on Ethereum and Binance Smart Chain, and the results suggest that {\framework} can raise timely warnings once contracts are deployed.
\end{abstract}

\input{sec-intro.tex}

\input{sec-background.tex}

\input{sec-threat.tex}

\input{sec-challenges.tex}

\input{sec-design.tex}

\input{sec-evaluation.tex}

\input{sec-conclusion.tex}



\bibliographystyle{plain}
\bibliography{ref.bib}

\end{document}

%% file: sec-intro.tex
\section{Introduction}
\label{sec:intro}
After Satoshi Nakamoto launched Bitcoin~\cite{nakamoto2008bitcoin}, blockchain platforms have sprung up. Among them, Ethereum~\cite{buterin2014next} is the most well-known one. 
Apart from these common characteristics among all blockchain platforms, the most eye-catching feature in Ethereum is the \textit{smart contract}. It can be taken as a piece of unchangeable script that shall be executed in a determined way once the pre-defined conditions are met.

As billions of USD equivalent assets are stored within smart contracts, identifying and exploiting hidden vulnerabilities in them is the top priority for attackers.
According to a recent report~\cite{slowmist2022securityreport}, there exist around 303 large attack events against well-known Ethereum contracts within the past year, amounting to losses of approximately \$3.8 billion.
Chen et al.~\cite{chen2020survey} have summarized 26 major types of vulnerabilities in Ethereum smart contracts, and new types of vulnerabilities are still emerging in an endless stream along the constant feature introduction and deprecation in Ethereum~\cite{kushwaha2022systematic}.

The automatic identification of vulnerabilities in Ethereum smart contracts is a well-studied topic~\cite{krupp2018teether, tikhomirov2018smartcheck, mythril2018, brent2020ethainter, ma2021security, kushwaha2022ethereum, ye2022vulpedia, grech2018madmax}.
Considering the number of contracts and the economic loss caused by false negatives, static symbolic execution, which can guarantee soundness to a certain extent, is used as the main analysis method.
For example, Mythril~\cite{mythril2018} is a static bytecode-level symbolic executor for Ethereum contracts, while Slither~\cite{feist2019slither} works on source code, which claims to achieve high efficiency and effectiveness.

Verifying the validity of the input addresses is a common practic and a key step to ensure the secure execution of smart contracts. Vulnerabilities in the process of address verification can lead to great security issues, and anecdotal evidence has been reported by our community~\cite{beosin2023visor}. 
To be specific, functions in smart contracts can be addressed as arguments. If developers accidentally neglect the verification on the passed address, once it is taken as the target of an external call, arbitrary operations, including malicious ones, within an address can be invoked. Therefore, if any on-chain state modification relies on that external call, such modification may go against the developer's wishes.

Currently, no existing tools can detect the address verification vulnerability, and implementing such a vulnerability detector based on existing framework is challenging.
On the one hand, according to the characteristics of the address verification vulnerability, to effectively identify it, the detector should perform an inter-procedural or even inter-contract analysis. Existing work, however, invariably suffers from either the efficiency or the effectiveness problem when conducting such analyses. For example, a pattern-based detector can barely handle such a complicated vulnerability pattern, while the symbolic execution suffers the path explosion issues and the bottleneck brought from the constraints solving.
On the one hand, most contracts are close-sourced, and bytecode lacks sufficient semantics, making it difficult to precisely identify hidden vulnerabilities. Specifically, the address verification vulnerability requires fine-grained tracking in both memory and storage areas. In bytecode, it is hard to distinguish different variables stored in these two areas. Moreover, tracking them via symbolic execution would suffer severe path explosion issue due to the large space of feasible slots.

\textbf{This work.}
To fill the void, we aim to characterize and detect this kind of emerging vulnerability. To be specific, we design and implement {\framework}, a novel lightweight static taint analysis framework that can efficiently and effectively identify the address verification vulnerability.
Its taint analysis relies on the static simulation of the opcode sequence. In other words, without considering the feasibility of paths, {\framework} maintains the values of data structures (e.g., stack and memory) and the taint propagation status.
Such a static simulation traverses lots of paths, including the originally infeasible ones.
Thus, according to the corresponding states collected, {\framework} formally detects the address verification vulnerability through a three-phase detection. It can progressively rule out false positives and false negatives based on the intrinsic characteristics of the vulnerability.

Based on our crafted benchmarks, {\framework} significantly outperforms state-of-the-art smart contract detection tools (i.e., Mythril, Ethainter, Jackal, and ETHBMC)\footnote{Note that these tools cannot detect address verification vulnerability, we have implemented the same detection logic atop them, for a fair comparison.} in both terms of efficiency and effectiveness. 
According to a comprehensive evaluation of over 5 million deployed Ethereum contracts, {\framework} flags over 812 vulnerable contracts, and 348 open-source smart contracts were further verified, whose total value locked is over \$11.2 billion.
Finally, we deploy {\framework} as a real-time detector on Ethereum and BSC, an EVM-like blockchain platform. The results show that {\framework} is capable of raising early warnings to developers and the community timely. A real-world case, where {\framework} raises the warning 1.5 hours ahead of the attack, proves its effectiveness and efficiency.

This paper makes the following contributions:

\begin{itemize}
\item This is the first work on detecting address verification vulnerability. We have designed and implemented {\framework}, an efficient and effective taint analyzer based on static EVM simulation.

\item We have applied {\framework} to over 5 million smart contracts on Ethereum, and uncovered hundreds of vulnerable smart contracts which were undisclosed by our community before. This is the first large scale in-the-wild characterization study of this kind of vulnerability.

\item Compared to Mythril, Ethainter, Jackal, and ETHBMC, {\framework} can improve the analysis efficiency around 2 to 5 times while achieving 94.3\% precision and 100\% recall on well-established benchmarks.

\item We have deployed {\framework} as a real-time detector on Ethereum and BSC. The results suggest that {\framework} is able to raise early warnings once contracts are deployed before the attack is initiated by attackers.
\end{itemize}

%% file: sec-background.tex
\section{Background}
\label{sec:background}

\subsection{Ethereum Primer}
\label{sec:background:contract}
\label{sec:background:defi}
There are two types of accounts in Ethereum: \textit{external owned account (EOA)} and \textit{smart contract}.
Specifically, an EOA is an ordinary account, which is identified by a unique address and controlled by a private key.
Furthermore, smart contracts can be regarded as scripts, which are mainly written in Solidity~\cite{dannen2017introducing}, a well-defined and easy-to-use programming language proposed by the Ethereum official.
Interactions among accounts are achieved by initiating transactions, which carry the corresponding data.
Smart contracts are executed in Ethereum Virtual Machine (EVM)~\cite{hirai2017defining}, which is embedded in Ethereum client nodes.
EVM is stack-based, and all data is stored either permanently or temporarily. Specifically, all operands of operators and intermediate values are pushed onto and popped from the \textit{stack}. The \textit{memory} area~\cite{hildenbrandt2018kevm}, the temporary one, only keeps data under the context of the current transaction. Only the data stored in the \textit{storage} area~\cite{ma2021security} is permanent, i.e., is stored on-chain.
We often denote a set of smart contracts which jointly achieve a specific functionality as a decentralized application (DApp).

DApps have demonstrated significant potential since their rise in 2016~\cite{he2020characterizing}. Many genres of DApps have emerged, e.g., gambling~\cite{norvill2017automated}, token swap~\cite{xia2021trade}, and lending~\cite{teng2022applications}.
Alongside the hundreds of billions of USD invested in Ethereum, the decentralized versions of traditional financial tools, e.g., exchanges and insurance, have appeared, which are called decentralized finance, i.e., DeFi.
Taking advantage of the decentralization, permissionlessness, and transparency in Ethereum, DeFi starts to rise like a rocket.
According to statistics, DeFi accounted for \$163 billion at the end of 2022~\cite{defillama2023tvl}.

Except for the official token, \texttt{Ether}, Ethereum allows users to issue tokens as they wish, as long as these tokens meet the standard like ERC-20~\cite{openzeppelin2023erc20}.
The ERC-20 standard consists of six mandatory functions. Any smart contract implements these functions can issue valid tokens that can circulate in Ethereum, like USDT~\cite{bullmann2019search} and USDC~\cite{arner2020stablecoins}.
Therefore, DeFi can also issue its own ERC-20 tokens, and take other ERC-20 tokens as valid ones. The interoperability between ERC-20 tokens and DeFi pushes the prosperity of Ethereum.

\subsection{Whitelisted Address Verification}
\label{sec:background:verification}
In Ethereum, examining the validity of the given addresses is a common practice, which is called \textit{whitelisted address verification}. It is widely adopted in DeFi apps such as lending~\cite{xu2022banks} and bank~\cite{alkhalifah2021mechanism}.
\textit{Address verification is the cornerstone to ensure the safety of smart contracts.}
Therefore, \texttt{OpenZeppelin}~\cite{openzeppelin2023access}, a well-known standard library provider in Ethereum, offers a whitelisted verification method. 
Moreover, through a comprehensive study of the top 40 DeFi projects ranked by TVL (Total Value Locked)~\cite{defillama2023ethereum}, which account for over 95\% of the total DeFi market, we summarized the verification techniques they adopted.
In short, three whitelisted verification methods are involved, i.e., \textit{hard-encoded comparison}, \textit{mapping validation}, and \textit{hard-encoded address enumeration}.
Note that, though we cannot guarantee all adopted address verification techniques are covered, we cover the most prevalent ones. 
Considering the extensive copy-and-use in Ethereum smart contracts~\cite{he2020characterizing}, these three mechanisms are representative.

\begin{listing}[t]
\begin{minted}{java}
function deposit(uint256 amount, address token) external {
    require(token == usdt, "not usdt token!");
    token.safeTransferFrom(msg.sender, address(this), amount);
}
\end{minted}
    \caption{An example of hard-encoded address comparison.}
    \label{list:Hardcoded Address Comparison}
\end{listing}

\begin{listing}[t]
\begin{minted}{java}
address[] public addresses;

function contains(address _address) public view returns (bool) {
    for(uint i = 0; i < addresses.length; i++) {
        if (addresses[i] == _address) {
            return true;
        }
    }
    return false;
}

function deposit(uint256 amount, address token) external {
    require(contains(token), "not contain this token");
    token.safeTransferFrom(msg.sender, address(this), amount);
}
\end{minted}
    \caption{An example of hard-encoded address enumeration.}
    \label{list:Whitelist Enumeration}
\end{listing}

Listing~\ref{list:Hardcoded Address Comparison} illustrates how \textbf{hard-encoded comparison} works. As we can see, the passed \texttt{token} at L2\footnote{L2 refers to the second line, we will adopt this notation in the paper.} is required to equal the address of \texttt{usdt}, otherwise it raises an exception.
\textbf{Mapping validation} adopts a \textit{mapping} structure that can dynamically maintain the whitelisted status of addresses, e.g., \texttt{mapping(address => bool) whitelist}.
As for the \textbf{hard-encoded addresses enumeration}, it is a variant of the first one. As shown in Listing~\ref{list:Whitelist Enumeration}, an array named \texttt{addresses} keeps all whitelisted addresses. Therefore, once the \texttt{deposit} function is invoked, the argument \texttt{token} is passed to the \texttt{contains} function, defined at L3, which is basically a hard-encoded comparison wrapped by a loop.
At the bytecode level, these three techniques perform similarly. The contract loads the address in arguments by \texttt{CALLDATALOAD} and performs examination via a conditional opcode \texttt{JUMPI}.
If an address is whitelisted, the control flow will be directed to the \textit{fallthrough} branch, and the following logic will be used. Otherwise, the \textit{jumpdest} branch is responsible for handling failed assertions.

\subsection{Taint Analysis on Smart Contracts}
\label{sec:background:taint}
Taint analysis is a fundamental method of program analysis, used for detecting vulnerabilities~\cite{frowis2019detecting} and tracking sensitive information flow~\cite{kushwaha2022ethereum}.
Before performing the taint analysis, \textit{sources} and \textit{sinks} should be specifically defined, where \textit{source} refers to input fields controlled by adversaries, and \textit{sink} refers to any part of the system where potentially dangerous data can be used in an unsafe manner.
Taint analysis will track data flow from sources to sinks and identify any operations or transformations on the data along the way.

In the context of Ethereum smart contracts, the source is often the function of smart contracts that accept transactions from other accounts, while the sink varies, depending on specific goals. For example, to examine if a contract can be destructed, Ethainter~\cite{brent2020ethainter} takes the \texttt{SELFDESTRUCT} opcode as the sink. Moreover, Michael et al.~\cite{frowis2019detecting} introduce a tool based on symbolic execution and taint analysis, designating \texttt{SSTORE} as the primary sink for its evaluations.

%% file: sec-threat.tex
\subsection{Threat Model}
\label{sec:threat}
Adversaries in our study do not require any extra privileges.
This is because Ethereum is a permissionless blockchain platform, which allows any non-privileged account, including malicious ones, to initiate transactions with sufficient gas, deploy valid smart contracts, and invoke any already deployed ones.
However, certain limitations still exist. For instance, they cannot breach the integrity of the Ethereum network or manipulate the block generation process, and cannot access the private keys of legitimate accounts.
In a nutshell, we can barely distinguish adversaries from well-behaved accounts.

%% file: sec-challenges.tex
\section{Motivating Example and Challenges}
\label{sec:challenges}

\begin{listing}[t]
\begin{minted}{java}
function deposit(
    uint256 visrDeposit,
    address payable from,
    address to
) external returns (uint256 shares) {
    require(visrDeposit > 0, "deposits must be nonzero");
    require(to != address(0) && to != address(this), "to");
    require(from != address(0) && from != address(this), "from");

    shares = visrDeposit;
    if (vvisr.totalSupply() != 0) {
        uint256 visrBalance = visr.balanceOf(address(this));
        shares = shares.mul(vvisr.totalSupply()).div(visrBalance);
    }

    if(isContract(from)) {
        require(IVisor(from).owner() == msg.sender); 
        IVisor(from).delegatedTransferERC20(address(visr), address(this), visrDeposit);
    }
    else {
        visr.safeTransferFrom(from, address(this), visrDeposit);
    }

    vvisr.mint(to, shares);
}
\end{minted}
    \caption{The vulnerable \texttt{deposit} function in \texttt{vvisr}.}
    \label{lst:vvisr}
\end{listing}

\subsection{Motivating Example}
\label{sec:challenges:motivating}
Listing~\ref{lst:vvisr} shows a smart contract owned by \texttt{Visor Finance} that is vulnerable to the address verification vulnerability. It was attacked on Dec. 21st, 2021~\cite{beosin2023visor}, causing \$8.2 million financial losses.
As we can see, the \texttt{deposit} function takes three arguments, namely, the number of tokens to be deposited (\texttt{visrDeposit}), the payer (\texttt{from}), and the beneficiary (\texttt{to}).
From L6 to L8, it performs sanity checks, i.e., a valid amount of deposit, and valid addresses for both payer and payee.
After that, it translates the deposit into shares according to \texttt{totalSupply()} (L11 to L14), performs the corresponding token transfer from the \texttt{from} address to itself (L16 to L22), and mints some \texttt{vvisr} tokens to the \texttt{to} address (L24).
The vulnerability is hidden in the if code block at L16.
Specifically, it allows the \texttt{from} address as a contract, and examines if its \texttt{owner} function returns the address of the transaction initiator (L17). If the assertion passes, then it invokes the \texttt{delegatedTransferERC20} function defined in \texttt{from}.
Recalling the threat model mentioned in \S\ref{sec:threat}, attackers are able to deploy contracts and initiate transactions arbitrarily. More specific, if the \texttt{from} is actually provided by some malicious ones, they can control the behaviors of L17 and L18.
To this end, the control flow will be successfully directed to L24, where \texttt{vvisr} finally issues tokens to \texttt{to}, which is also controlled by attackers, without receiving any tokens from \texttt{from} that is expected by developers of \texttt{Visor Finance}.

Through this example, we can summarize three principles related to the address verification vulnerability:
\begin{itemize}
	\item[\textbf{P1}] \textit{The vulnerable function takes an address as a parameter, and performs insufficient authorization examination on that address.} Through the address, attackers can pass self-deployed and unauthorized contracts.
	\item[\textbf{P2}] \textit{The address in \textbf{P1} is taken as the target of an external call}. Through the external call, the control flow is transferred to attackers. Thus, they can totally control the behavior of the external call, including the return value.
	\item[\textbf{P3}] \textit{On-chain states that are control-flow dependent on the return value mentioned in \textbf{P2} are updated}. To this end, through an unauthorized control flow, adversaries can get profits by indirectly manipulating on-chain states, like initiating an external call or updating balance.

\end{itemize}

\subsection{Challenges}
\label{sec:challenges:challenges}
In response to the address verification vulnerability, as outlined in the summarized principles and the motivating example in \S\ref{sec:challenges:motivating}, we identify the following challenges.

\noindent
\textit{\textbf{C1: Lack of semantics.}}
It is challenging to precisely identify if the address mentioned in \textbf{P1} is sufficiently verified due to the lack of semantics in bytecode.
According to the statistics~\cite{ma2023abusing}, more than 99\% of Ethereum contracts have not released their source code.
The bytecode format is quite unreadable and contains little semantic information.
Moreover, there is no debug information to assist in recovering the semantics.
Traditional bytecode-based analysis tools usually require some methods to overcome this challenge, like symbolic execution~\cite{mythril2018}.

\noindent
\textit{\textbf{C2: Inter-procedural analysis on control flow and data flow.}}
Detecting this vulnerability requires accurately extracting the control flow and data flow dependencies inter-procedurally.
Specifically, in \textbf{P2}, there is an external call to an address, which is passed via the argument. Between the external call and the function entry, it will be propagated several times due to the authorization verification. Thus, we have to precisely identify if the callee address is one of the arguments through parsing data flow.
Moreover, in \textbf{P3}, the on-chain state update depends on the return value of the external call in \textbf{P2} in terms of control flow, which requires us to identify control flow dependencies among variables.
In addition, from Listing~\ref{list:Whitelist Enumeration} we can conclude that some authorizations are verified in other functions, which requires inter-procedural analysis.

\subsection{Limitations of Existing Tools}
\label{sec:challenges:limitations}
Considering the aim of implementing a lightweight and effective detector for address verification vulnerability, we exclude the analyzers that adopt dynamic analysis.
The reasons for this decision are twofold.
On the one hand, dynamic analysis requires a runtime environment for execution,  which is resource-intensive and time-consuming, contrary to our lightweight goal.
On the other hand, dynamic analysis identifies vulnerabilities with generated test cases and oracles. For intricate contracts, especially for inter-contract analysis, this method might not always cover all vulnerable paths, potentially resulting in false negatives.
Consequently, static methods are considered, including \textit{pattern-based matching}, \textit{symbolic execution}, and \textit{taint analysis}.
\textit{To the best of our knowledge, no existing tools can be directly deployed to detect this vulnerability}. Although we can extend them with the ability to detect the vulnerability, we observe some intrinsic limitations.

\noindent
\textbf{Pattern-based Matching.} It relies on heuristic rules summarized by developers. Lots of preliminary tools adopt this method, identifying vulnerabilities according to opcode sequences~\cite{grech2018madmax}, transaction histories~\cite{ali2021sescon}, and call traces~\cite{chen2019dataether}.
However, such a manual process is unsound and error-prone, which cannot even handle \textbf{C1}, i.e., identifying semantics of bytecode sequences.
Consequently, the constant update of Solidity syntax~\cite{wohrer2018smart} and compilation toolchain~\cite{remix2020} make this kind of analyzers ineffective.

\noindent
\textbf{Symbolic Execution \& Model Checking.}
Both techniques are widely adopted in identifying vulnerabilities in software analysis.
Taking advantages of abstraction on programs, symbolic executors and model checkers can recover some semantic information to overcome \textbf{C1}.
However, they are inherently limited by the efficiency issue, mainly brought by the path/state explosion.
For example, when dealing with the example shown in Listing~\ref{lst:vvisr}, both of them should not only traverse all possible functions through the entry to find the \texttt{deposit}-like functions, but also get stuck from L16 to L18, where they would try all possible contracts.
Because they cannot effectively and efficiently conduct the inter-contract or even inter-procedural analysis, \textbf{C2} is the main challenge hiders the adoption of them.
Moreover, they try to model the memory area in a precise way, i.e., considering every bit in the memory. Although such a precise modeling benefits to the soundness of analyzing results, it comes at the cost of efficiency. In the context of address verification vulnerability, we only need to focus on the propagation of address parameters in the memory area instead of concrete data.

\noindent
\textbf{Taint Analysis.} Existing taint analyzers cannot be directly adopted to identify this vulnerability.
Firstly, some of them rely on the source code~\cite{feist2019slither}. In Ethereum, such contracts only account for less than 1\%. 
Secondly, some of them are limited by their adopted methods when collecting the taint propagation information and tracking taint propagation.
For instance, Sereum~\cite{rodler2018sereum} uses dynamic analysis to monitor execution at runtime. Furthermore, Mythril~\cite{mythril2018} and Osiris~\cite{torres2018osiris} both adopt symbolic execution to statically collect information. Mythril tries to traverse all feasible paths, while Osiris is only intra-procedural.
Ethainter~\cite{brent2020ethainter} is a strong contender.
However, it exhibits proficiency in addressing \textbf{C1} and \textbf{C2}, but it suffers due to the intricacies of EVM’s linear memory model. Specifically, it cannot properly handle dynamic memory allocations and deallocations in EVM's memory, which potentially compromises the accuracy of its taint tracking, especially when contracts execute complex memory operations or utilize memory for storing and manipulating intermediate data (like the three memory verification mechanisms mentioned in \S\ref{sec:background:verification}).

\noindent
\textbf{\textit{Our Key Idea.}}
As aforementioned, symbolic execution and taint analysis can effectively recover the semantics and conduct the inter-procedural analysis on control flow and data flow dependencies to some extent.
However, considering the soundness of the analysis, symbolic execution always suffers the efficiency issue.
To this end, we decided to adopt \textit{static EVM simulation based taint analysis}, which tracks the taint propagation through a static simulation on the bytecode.
To reduce the false positives introduced by static simulation and filter out paths that can lead to exploitation, we design a three-stage detector corresponding to the three principles mentioned in \S\ref{sec:challenges:motivating}.
Moreover, to overcome the memory issue suffered by Ethainter, we decide to design the memory model motivated by He et al.~\cite{he2021eosafe} to abstract the sparse linear memory into key-value pairs.
Consequently, our design can address both \textbf{C1} and \textbf{C2}.
Furthermore, {\framework} simulates stack and memory through a self-implemented EVM simulator to capture the propagation of addresses, ensuring efficiency for non-dynamic data. As for the dynamic memory parameters, {\framework} conservatively treats them as symbols and only guarantees the stack balance to avoid the negative influence brought by enumerating all possible values.

%% file: sec-design.tex
\section{Design of {\framework}}
\label{sec:design}

This section elucidates the technical intricacies of {\framework}, which is designed to detect the address verification vulnerability in Ethereum smart contracts. We firstly give a high-level overview in \S\ref{sec:design:overview}, and an introduction of adopted notations in \S\ref{sec:design:notations}. Then, we delve into the three components, respectively, from \S\ref{sec:design:m1} to \S\ref{sec:design:m3}.

\begin{figure*}
    \centering
    \includegraphics[width=0.8\linewidth]{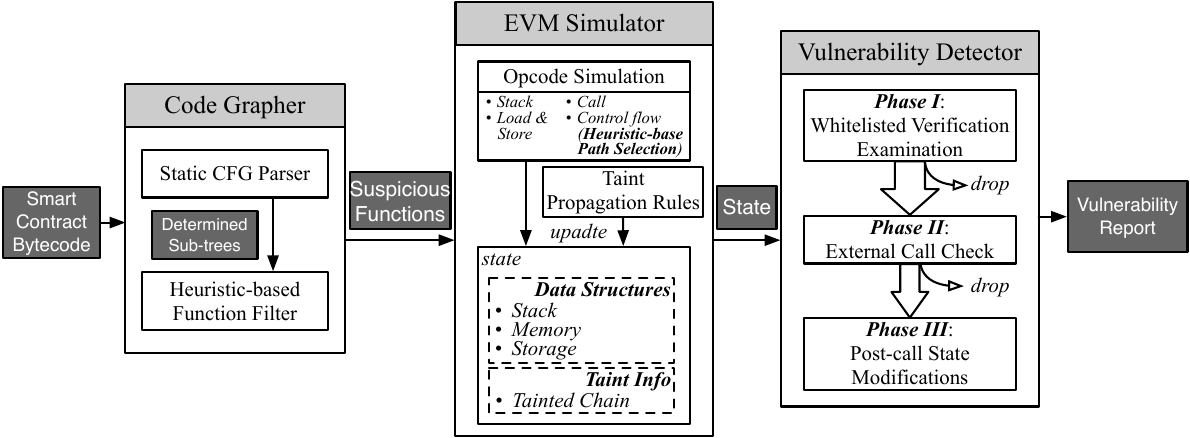}
    \caption{The workflow and architecture of {\framework}.}
    \label{fig:overview}
\end{figure*}

\subsection{Overview}
\label{sec:design:overview}
Fig.~\ref{fig:overview} illustrates the architecture and workflow of {\framework}, which is composed of three main components, i.e., \textit{code grapher} (denoted as \textit{{\grapher}}), \textit{EVM simulator} (denoted as \textit{{\simulator}}), and \textit{vulnerability detector} (denoted as \textit{{\detector}}).
Specifically, {\framework} only takes the bytecode of a Solidity smart contract as input. The {\grapher} parses it into the control flow graph (CFG), filters out all suspicious functions as candidates, and delivers them to the {\simulator}.
The {\simulator} maintains a state, consisting of two parts.
One part is the data structures required by EVM, i.e., stack, memory, and storage (see \S\ref{sec:background:contract}); the other part is the collected taint information. According to the CFG, the {\simulator} updates fields in states according to the opcode sequence. It also adopts a heuristic-based path selection method to focus on the most valuable path, i.e., the ones that may lead to the vulnerability.
Once the analysis against a path is finished, the corresponding state is sent to the {\detector} to determine if the current contract is vulnerable to the address verification vulnerability.
The cascaded three-phase detection strategy in the {\detector} rules out false positives and false negatives based on the intrinsic characteristics (\textbf{P1} to \textbf{P3}).

\begin{figure}
    \centering
    \includegraphics[width=0.9\linewidth]{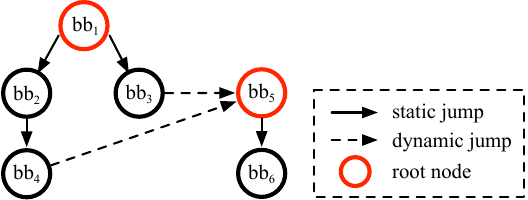}
    \caption{The CFG of \texttt{foo}.}
    \label{fig:Partial Control Flow Graph}
\end{figure}

\subsection{Notations}
\label{sec:design:notations}

To better explain the implementation of {\framework}, we define some notations here:

\begin{itemize}
\item $\mathcal{S}$, the set of sources that can be controlled by users;
\item $\mathcal{T}$, the set of tainted variables;
\item $C_{\mathcal{T}}$, a mapping from a tainted variable to its sources;
\item $\mathcal{F}$, the set of suspicious functions;
\item $Mem$, $Sto$, refer to the memory and storage area in EVM.
\item $V$, $EC$, and $SM$, refer to the three-stage detection in the {\detector}, respectively. Each of them takes a function $f$ and a parameter $p$ as inputs.
\end{itemize}

\subsection{Component\#1: Code Grapher}
\label{sec:design:m1}
Generally speaking, the {\grapher} is responsible for obtaining the sub-tree of CFG of the given function.
Given a piece of bytecode, {\grapher} firstly extracts the runtime code, consisting of implementation of functions~\cite{he2020characterizing}.
Then, {\grapher} parses it into basic blocks and constructs the CFG according to their jump relations.
However, some jump relations are determined dynamically at runtime instead of statically at the compilation stage.
Thus, the {\grapher} constructs the CFG only on statically determined jump relations considering the soundness.
Take the Fig.~\ref{fig:Partial Control Flow Graph} as an example, where $bb_1$ is the entry of the function \texttt{foo} and its jump relations to $bb_2$ and $bb_3$ can be statically determined.
Though $bb_3$ and $bb_4$ can jump to $bb_5$ at runtime, they are determined dynamically. 
Thus, though the {\grapher} generates two trees whose roots are $bb_1$ and $bb_5$, respectively, $bb_5$ is actually a subtree of $bb_1$ at runtime.

To filter out suspicious functions, i.e., the ones that may be vulnerable to the address verification vulnerability, the {\grapher} heuristically keeps functions that take addresses as arguments (\textbf{P1}).
Specifically, each address parameter undergoes a bitwise AND operation with \texttt{0xFF..FF} (160-bit long). By identifying such a specific pattern, which is also widely adopted in previous work~\cite{amani2018towards}, we can extract functions that meet \textbf{P1}. These functions will be added to the set $\mathcal{F}$.

\subsection{Component\#2: EVM Simulator}
\label{sec:design:m2}
Based upon the $\mathcal{F}$ passed by the {\grapher}, the {\simulator} is responsible for: 1) maintaining the data structures required by EVM, and 2) updating the taint information.
Specifically, $\mathcal{S}$ is composed of all user-controllable variables, e.g., \texttt{ORIGIN}, \texttt{CALLDATA}, and \texttt{BALANCE}.
For each opcode, the {\simulator} specifies a set of rules to update both data structures and taint information in a static way.
Without loss of generality, EVM opcodes are classified into four categories: \textit{stack related}, \textit{load \& store related}, \textit{call related}, and \textit{control flow related}.

\subsubsection{Stack Related Opcodes}
The stack-related opcodes include the ones that only interact with the stack and do not change the control flow, like \texttt{ADD}, \texttt{SWAP}, and \texttt{DUP}.
Take an opcode that takes two arguments ($op_{\text{uid}_1}, op_{\text{uid}_2}$) and returns one ($op_{\text{uid}_k}$) as an example, like \texttt{ADD}.
Except for conducting necessary update on the stack, we also formalize the taint propagation rules as follows:

\begin{equation}
\begin{gathered}
    \{op_{\text{uid}_1}, op_{\text{uid}_2}\} \cap (\mathcal{S} \cup \mathcal{T}) \neq \emptyset\ \vdash\ \mathcal{T} \coloneqq \mathcal{T} \cup \{op_{\text{uid}_k}\}\\
    C_{\mathcal{T}}[op_{\text{uid}_k}] \coloneqq \begin{cases}
    \{op_{\text{uid}_1}\}, & op_{\text{uid}_1} \in \mathcal{S} \cup \mathcal{T}.\\
    \{op_{\text{uid}_2}\}, & op_{\text{uid}_2} \in \mathcal{S} \cup \mathcal{T}. \\
    \{op_{\text{uid}_1}, op_{\text{uid}_2}\}, & \{op_{\text{uid}_1}, op_{\text{uid}_2}\} \cap (\mathcal{S} \cup \mathcal{T}) \neq \emptyset.
    \end{cases}
\end{gathered}\nonumber
\end{equation}

The first rule implies that if any operand belongs to $\mathcal{S}$ or $\mathcal{T}$, the return value shall be added to the $\mathcal{T}$, i.e., tainted set.
The second rule updates $C_{\mathcal{T}}$, which keeps the tainted relation from $op_{\text{uid}_k}$ to its direct taint predecessor.
Note that, for other opcodes under this category but with different numbers of arguments and return value, the propagation rules are similar.

\subsubsection{Load \& Store Related Opcodes}
As we mentioned in \S\ref{sec:background:contract}, memory and storage are maintained by EVM. Data in them are organized in different ways.
Specifically, storage operates like a key-value dictionary, i.e., the values are retrieved by the given key~\cite{albert2022max}.
As for the memory, it can be seen as a flat and contiguous array of bytes. Accessing to an item in memory is achieved by a calculated offset.
However, in Ethereum smart contracts, a memory is always sparse. Thus, we are motivated by the method raised by He et al.~\cite{he2021eosafe}, i.e., abstracting the memory area as a key-value pairs, where key is the offset and value is the corresponding data.
To this end, data in memory and offset can be retrieved and indexed uniformly, like $Sto[key]$ or $Mem[offset]$.

\noindent
\textbf{Load Related Opcodes.}
This includes \texttt{MLOAD} and \texttt{SLOAD}, i.e., load from memory or storage, respectively. Both of them take a single argument, $tar$, i.e., the target address, and return $ret$, indicating the retrieved data.
The taint rules are defined as:

\begin{equation}
\begin{gathered}
    tar \in \mathcal{S} \cup \mathcal{T}\ \vdash\ \mathcal{T} \coloneqq \mathcal{T} \cup \{ret\} \\
    C_{\mathcal{T}}[ret] \coloneqq \{tar\},\ \text{if}\ tar \in \mathcal{S} \cup \mathcal{T}
\end{gathered}\nonumber
\end{equation}
In other words, if $tar$ is tainted, $ret$ will be marked as tainted. Their dependency relation will be kept in $C_{\mathcal{T}}$.
Moreover, there is another situation, i.e., the data designated by $tar$ is tainted already. In this way, we set the taint propagation rules as (take retrieving from memory area as an example):

\begin{equation}
\begin{gathered}
    Mem[tar] \in \mathcal{S} \cup \mathcal{T}\ \vdash\ \mathcal{T} \coloneqq \mathcal{T} \cup \{ret\} \\
    C_{\mathcal{T}}[ret] \coloneqq \{Mem[tar]\},\ \text{if}\ Mem[tar] \in \mathcal{S} \cup \mathcal{T}
\end{gathered}\nonumber
\end{equation}

\noindent
\textbf{Store Related Opcodes.}
Both \texttt{SSTORE} and \texttt{MSTORE} take $val$ and $dest$ as arguments, referring to the to-be-stored data and the destination, respectively.
Thus, we can formalize the taint propagation rules as:
\begin{equation}
\begin{gathered}
    \{val, dest\} \cap (\mathcal{S} \cup \mathcal{T}) \neq \emptyset\ \vdash\ \mathcal{T} \coloneqq \mathcal{T} \cup \{val\} \\
    C_{\mathcal{T}}[val] \coloneqq \{dest\},\ \text{if}\ dest \in \mathcal{S} \cup \mathcal{T}
\end{gathered}\nonumber
\end{equation}

That is to say, if any of the $val$ or $dest$ is tainted, the final stored $val$ will be tainted. When maintaining the taint dependency relation, we only consider if $val$ depends on $dest$. There does not exist an edge from $val$ to $val$.

\subsubsection{Call Related Opcodes}
Several opcodes can conduct an external call, e.g., \texttt{CALL}, \texttt{DELEGATECALL}, and \texttt{SELFDESTRUCT}.
Among all arguments, the address is what we concerned, denoted as $param_{addr}$.
Therefore, we customize the taint propagation rules as:
\begin{equation}
\begin{gathered}
    C_{\mathcal{T}}.ancestor(param_{addr}) = \texttt{CALLDATALOAD}\ \vdash\ \mathcal{T} \coloneqq \mathcal{T} \cup \{suc\} \\
    C_{\mathcal{T}}[suc] \coloneqq \{param_{addr}\}
\end{gathered}\nonumber
\end{equation}

Specifically, our main concern is whether the ancestor of $param_{addr}$ can be tainted from \texttt{CALLDATALOAD}, which is the only source in $\mathcal{S}$ that can parse the address-typed variables. Thus, if $ C_{\mathcal{T}}.ancestor(param_{addr}) = \texttt{CALLDATALOAD}$ holds, it means that attackers can deliver a malicious address through the function entry to that external call finally.

\subsubsection{Control Flow Related Opcodes}
\label{sec:design:m2:control}
According to the specification, some of them do not take arguments, like \texttt{RETURN}, \texttt{STOP}, \texttt{REVERT}, and \texttt{INVALID}. Therefore, no taint marks are propagated when simulating these instructions. The {\simulator} only performs the control flow simulation on them.
For example, when the {\simulator} encounters the \texttt{INVALID} instruction, it halts the simulation of the current path and moves to the next one.

\texttt{JUMP} and \texttt{JUMPI} opcodes are crucial in handling taint propagation as they take the jump destination to explicitly alter the control flow, where \texttt{JUMPI} takes an argument as the condition. Because \texttt{JUMP} can be seen as a special edition of \texttt{JUMPI}, we demonstrate the handling on \texttt{JUMPI} in the following.
Specifically, \texttt{JUMPI} takes two arguments, i.e., $dest$ and $cond$, referring to the jump destination and the jump condition, respectively. It returns no value and directly performs the control flow jump.
For each \texttt{JUMPI}, there are two following branches, \textit{fallthrough} and \textit{jumpdest}.
The former one corresponds to the successive opcode, which is executed once the condition is not met, while the latter one is executed when the condition meets.
According to the specification~\cite{rodler2018sereum}, $dest$ should not rely on any arguments, i.e., determined statically during the compilation. Therefore, according to whether the $cond$ is determined dynamically, the {\simulator} adopts the following rules:

\begin{enumerate}
\item If the $cond$ is a concrete number, either \textit{fallthrough} or \textit{jumpdest} is chosen by the {\simulator} deterministically.
\item Otherwise, two paths are all feasible.
    \begin{enumerate}
    \item If the ancestor of the $cond$ is \texttt{CALLDATALOAD}, prioritize the \textit{jumpdest} path, i.e., \(cond = \text{True}\), to emulate attackers have successfully bypassed the examination.
    \item Otherwise, it takes two paths into consideration.
    \end{enumerate}
\end{enumerate}

The branch prioritization in step 2(a) is heuristic.
Specifically, once a condition is tainted from a source, it means that it is totally \textit{controllable} for attackers, like L17 in Listing~\ref{lst:vvisr}, i.e., attackers can always bypass the assertion by constructing the \texttt{owner} function.
On bytecode level, there is a conditional jump at L17, one is to L18 associated with a met condition, another is not shown in Listing~\ref{lst:vvisr}, indicating the \texttt{require} at L17 fails. Thus, the heuristic in step 2(a) prioritizes the branch to L18, i.e., attackers successfully bypass verifications.
Compared to symbolic execution, which collects path conditions along the simulation and queries the back-end SMT solver to obtain a concrete set of solutions, the advantages of this heuristic are twofold.
On the one hand, collecting path conditions and asking for solving is time- and resource-consuming, which is the greatest bottleneck~\cite{contro2021ethersolve}.
On the other hand, symbolic execution sometimes even cannot handle complex address verification logic. For example, when handling the L17 in Listing~\ref{lst:vvisr}, the symbolic executor has to conduct an inter-contract analysis to \texttt{IVisor} to obtain the value of \texttt{owner()}, which cannot be completed within an acceptable time.
Though step 2(b) may introduce some false positives, it is a tradeoff between the efficiency and effectiveness. In \S\ref{sec:evaluation:rq1}, it proves its effectiveness and practicability.

As for the taint propagation rules, we should note that no values are returned by these opcodes. Moreover, they have no side effects on both memory and storage.
In summary, no taint marks are propagated through these opcodes.

\subsection{Component\#3: Vulnerability Detector}
\label{sec:design:m3}

Based on the information collected from the {\simulator}, i.e., $\mathcal{F}$ and $C_{\mathcal{T}}$,
the {\detector} is able to determine whether a contract is vulnerable.
Specifically, as Fig.~\ref{fig:overview} illustrates, the risk detector is composed of three sequential phases, corresponding to the three principles mentioned in \S\ref{sec:challenges:motivating} (\textbf{P1} to \textbf{P3}).
We detail these three phases in the following.

\subsubsection{Phase I: Whitelisted Verification Examination}
\label{sec:design:m3:phasei}
The first phase is to examine whether a contract adopts a whitelisted verification method.
Therefore, we design a function $V(f_i, param_j)$. The function takes the $i$-th function from $\mathcal{F}$ and its $j$-th parameter as input, verifies if the parameter is examined by any of the methods described in \S\ref{sec:background:verification}.
The logic of $V(f_i, param_j)$ is defined as:
\begin{enumerate}
\item If $param_j$ is not typed as an address, we do not regard this situation to be valid. $V(f_i, param_j)$ returns \textit{True};
\item Against each \texttt{JUMPI} instruction, we examine if its $cond$ is tainted by the $param_j$, where $param_j$ is further tainted by the \texttt{CALLDATALOAD}. Formally, it is behaved like $param_j \in C_{\mathcal{T}}[cond] \land \texttt{CALLDATALOAD} \in C_{\mathcal{T}}[param_j]$. If it is hold, $V(f_i, param_j)$ returns \textit{True};
\item Otherwise, the tuple $(f_i, param_j)$ returns \textit{False}.
\end{enumerate}

Note that, the first two steps return \textit{True}, indicating a whitelisted verification is inapplicable or conducted normally. In other words, only the states with the \textit{False} return are kept and sent to the phase II check.

\subsubsection{Phase II: External Call Check}
According to \textbf{P2}, in the second phase, we further investigate if the $param_j$ can be used as the target of any external call related instructions. Therefore, we implement a function, named $EC(f_i, param_j)$
, which returns \textit{True} if the parameter is used as the target of an external call instruction in the function.

In Ethereum, an external call takes an address as a parameter, and allows the current contract to interact with them.
The implementation of $EC(f_i, param_j)$ is also intuitive. Specifically, if the tainted predecessor of the \textit{tar} of an external call instruction is $param_j$, or the $param_j$ itself, $EC(f_i, param_j)$ returns \textit{True}. Otherwise, it returns \textit{False}:
\begin{align}
EC(f, p) := \begin{cases}
    True, & p \in C_{\mathcal{T}}[tar] \vee tar = p.\\
    False, & \text{otherwise}.
    \end{cases}\nonumber
\end{align}

Similarly, to avoid meaningless resource consumption, only the states that correspond to the \textit{True} return value are passed to the third phase check. We regard the ones with \textit{False} return value as worthless vulnerable contracts.

\subsubsection{Phase III: Post-call State Modifications}
According to \textbf{P3}, we concern whether on-chain states can be updated according to the return value of the external call instructions focused by the phase II detection.
Therefore, we implement $SM(f_i, param_j)$.
In Ethereum, on-chain state update can be achieved by two ways.
On the one hand, some on-chain states can be seen as ordinary variables and modified directly, e.g., \texttt{BALANCE}. On the other hand, \texttt{SSTORE} instructions can also be used to alter on-chain state (see \S\ref{sec:background:contract}).
Therefore, for all these valuable targets, i.e., on-chain states keywords and \textit{val} and \textit{dest} of \texttt{SSTORE}, we put them in a set $\mathcal{V}$.
$SM$ returns \textit{True} if any element in $\mathcal{V}$ has a tainted predecessor that is related to the return value, i.e., $suc$, of the external call concerned by the phase II.
Formally,
\begin{align}
SM(f, p) := 
\begin{cases}
True, & \exists e \in \mathcal{V}.\  suc \in C_{\mathcal{T}}[e] \\
False, & \text{otherwise}.
\end{cases}\nonumber
\end{align}

Consequently, for all states with returned value as \textit{True}, they are kept as final ones.

\subsubsection{Address Verification Vulnerable Contracts}
In a nutshell, through such a three-phase detection, the {\detector} can effectively identify a state that can be exploited due to the existence of address verification vulnerability.
We can formally summarize our detection strategy as follows.

By parsing states passed from {\simulator}, {\detector} can obtain a set of tuples, consisting of potential victims:
\begin{equation}
\begin{gathered}
	Tuples = \{(f_i, param_j),\ \ldots\} := parse(states)
\end{gathered}\nonumber
\end{equation}

Through a three-phase detection, only the valuable and vulnerable states are remained:
\begin{equation}
\begin{gathered}
	Remained = \{(f, p) \in Tuples\ |\\\ \neg V(f, p) \land EC(f, p) \land SM(f, p)\}
\end{gathered}\nonumber
\end{equation}

If a contract has a state that is corresponded to a tuple in $Remained$, the contract is vulnerable to the address verification vulnerability.

%% file: sec-evaluation.tex
\section{Evaluation}
\label{sec:evaluation}

\subsection{Experimental Setup \& Research Questions}
\label{sec:evaluation:rq}
\textbf{Baselines.} 
To the best of our knowledge, no existing smart contract analysis framework supports the detection of address verification vulnerability.
However, to illustrate the effectiveness of {\framework}, we have extended 
Mythril~\cite{mythril2018} (commit: \texttt{f5e2784}), Ethainter~\cite{brent2020ethainter}, Jackal~\cite{gritti2023confusum} (commit: \texttt{3993e5c}), and ETHBMC~\cite{frank2020ethbmc} (commit: \texttt{e887f33}), four popular contract analyzers, as baselines.
Specifically, Mythril is a well-known and widely adopted symbolic executor that is specifically designed to detect vulnerabilities in Ethereum smart contracts.
Thus, we firstly integrate the same taint propagation rules into Mythril. Then, we employ the same three-stage detection logic as we stated in \S\ref{sec:design:m3}.
Consequently, Mythril is modified as a static symbolic execution based taint analyzer.
As for Ethainter, it is a source-code level reputable, extensively utilized, and scalable taint analyzer based on Datalog.
After being given a source code, Ethainter will first perform a complete analysis and extract the control flow and data flow dependencies in the contract. Therefore, we also implement a three-stage detector in Datalog to filter out the ones that comply with the rules mentioned in \S\ref{sec:design:m3} from all generated states.
Regarding Jackal, it decompiles Ethereum smart contracts' bytecode to an intermediate representation for constructing control flow graphs. We integrate the same taint propagation rules in Jackal, and employ the three-stage detection logic outlined in \S\ref{sec:design:m3}. This modification enhances Jackal’s capability to focus on identifying address verification vulnerabilities in smart contracts.
Last, ETHBMC utilizes bounded model checking and symbolic execution. We also manually integrate the three-stage detection logic into it to sharpen its focus on identifying address verification vulnerabilities.
Because detectors in all these tools follow the same set of principles on semantic level, this can reflect the distinction among them in terms of effectiveness and efficiency.

\noindent
\textbf{Implementation of {\framework}.}
{\framework} is written in Python and consists of 1.3K lines of code. As shown in Fig.~\ref{fig:overview}, it is composed of three main modules:

\textit{Code Grapher.}
It is responsible for disassembling the given bytecode into opcodes, and constructing the CFG according to the statical function call opcodes as we mentioned in \S\ref{sec:design:m1}.
Furthermore, in the function selector, the {\grapher} selects suspicious functions according to the heuristic in \S\ref{sec:design:m1}, and obtains the corresponding entry basic block.

\textit{EVM Simulator.}
The body of the {\simulator} is basically a two-layer nested loop, the outer one iterates all suspicious functions collected from the {\grapher}, and the inner one iterates all opcodes.
According to the taint propagation rules defined in \S\ref{sec:design:m2}, a \textit{state}, which is composed of data structures like EVM stack, memory, storage, and taint information, is updated.
When \texttt{JUMPI} is encountered, {\framework} employs the heuristic-based path selection approach.
The final state of each path will be sent to the {\detector}. Once a function is labelled as vulnerable, the inner loop will break to the next function to improve the analysis efficiency.

\textit{Vulnerability Detector.}
Leveraging the state yielded by the {\simulator}, all states undergo a three-phase check as introduced in \S\ref{sec:design:m3}.
If any state can pass all three phases, i.e., vulnerable, it will be returned to {\simulator} immediately.

\noindent \textbf{Experimental Setup.}
The experiments are conducted on a 48 core server equipped with two Intel Xeon 6248R processors, accompanied by 256GB RAM, while its time limitation for each contract is 10 minutes.

We answer the following research questions (RQs):

\begin{itemize}
\item[\textbf{RQ1}] Is {\framework} efficient and effective in identifying the address verification vulnerability?
\item[\textbf{RQ2}] How many smart contracts are vulnerable in the wild and 
what are their characteristics? 
\item[\textbf{RQ3}] Can {\framework} be deployed as a real-time detection system?
\end{itemize}

To answer RQ1, we launch {\framework} and four baseline tools on a well-constructed benchmark and real-world contracts in the wild, ranging from the genesis block to the one with the height of 17,421,000, created at Jun. 6th, 2023, around 59 million contracts in total.
Among these contracts, we regard the ones that are involved in at least one piece of transaction as the worth being analyzed one. Thus, 5,158,101 smart contracts remain as candidates.
To answer RQ2, we characterize the vulnerable contracts from different perspectives, e.g., number of transaction and tokens involved.
To answer RQ3, we have deployed {\framework} on both Ethereum and BSC~\cite{bnbchain2023}, a well-known EVM-like blockchain platform with the market cap as \$37.7 billion~\cite{coinmarketcap2023bnb}.
We comprehensively its usability and scalability of being a real-time detector.

\subsection{RQ1: Effectiveness \& Efficiency}
\label{sec:evaluation:rq1}

\subsubsection{Evaluating Results on Benchmark}
\label{sec:evaluation:rq1:benchmark}

\begin{table*}[]
\centering
\caption{Performance comparison among {\framework}, Mythril, Ethainter, Jackal, and ETHBMC on the benchmark.}
\label{table:benchmark}
\resizebox{\textwidth}{!}{%
\begin{tabular}{@{}ccccc|cccc|cccc|cccc|ccccc@{}}
\toprule
\multirow{2}{*}{\textbf{Metrics}} & \multicolumn{4}{c}{\textbf{{\framework}}} & \multicolumn{4}{c}{\textbf{Mythril}*} & \multicolumn{4}{c}{\textbf{Ethainter}*}  & \multicolumn{4}{c}{\textbf{Jackal}*} & \multicolumn{4}{c}{\textbf{ETHBMC}*}\\ \cmidrule(l){2-21} 
                                  & \textbf{P} & \textbf{$\overline{\textbf{P}}$} & \textbf{N} & \textbf{$\overline{\textbf{N}}$} & \textbf{P} & \textbf{$\overline{\textbf{P}}$} & \textbf{N} & \textbf{$\overline{\textbf{N}}$} & \textbf{P} & \textbf{$\overline{\textbf{P}}$} & \textbf{N} & \textbf{$\overline{\textbf{N}}$} & \textbf{P} & \textbf{$\overline{\textbf{P}}$} & \textbf{N} & \textbf{$\overline{\textbf{N}}$} & \textbf{P} & \textbf{$\overline{\textbf{P}}$} & \textbf{N} & \textbf{$\overline{\textbf{N}}$}\\ \midrule
Avg. Time (s)                     & 7.98       & 6.85                                  & 6.74           & 7.72                                 & 24.36           & 31.62                                 & 29.39      & 28.47                                 & 9.74 & 10.32 & 12.36 & 12.15  & 20.21 & 18.40 & 18.05 & 20.04  & 0.33 & 0.35 & 1.64 & 1.52\\
\rowcolor[HTML]{C0C0C0}
\# Timeout                        & 0          & 0                                 & 0           & 0                                 & 2           & 2                                  & 2          & 2                                 & 1 & 1 & 1 & 1   & 2 & 2 & 1 & 1  & 0 & 0 & 0 & 0\\
True Positives                    & 6          & -                                 & -           & 4                                 & 2           & -                                 & -          & 1                                & 4 & - & - & 3        & 4 & - & - & 3        & 0 & - & - & 0\\
True Negatives                    & -          & 6                                 & 4           & -                                & -           & 4                                 & 2          & -                                & - & 5 & 3 & -        & - & 4 & 3 & -        & - & 6 & 4 & -\\
\rowcolor[HTML]{C0C0C0}
False Positives                   & -          & 0                                 & 0           & -                                 & -           & 0                                 & 0          & -                                & - & 0 & 0 & -      & - & 0 & 0 & -        & - & 0 & 0 & -\\
\rowcolor[HTML]{C0C0C0}
False Negatives                   & 0          & -                                & -           & 0                                 & 2           & -                                 & -          & 1                                & 1 & - & - & 0      & 0 & - & - & 0       & 6 & - & - & 4\\ 
Precision & \multicolumn{4}{c|}{100\%} & \multicolumn{4}{c|}{100\%} & \multicolumn{4}{c|}{100\%} & \multicolumn{4}{c|}{100\%} & \multicolumn{4}{c}{0\%} \\
Recall                            & \multicolumn{4}{c|}{100\%} & \multicolumn{4}{c|}{50\%} & \multicolumn{4}{c|}{87.5\%} & \multicolumn{4}{c|}{100\%} & \multicolumn{4}{c}{0\%} \\
\bottomrule
\multicolumn{13}{l}{*The address vulnerability detector is implemented by ourselves.}                                                                                                                                                                                                                                                                                                                                                                                
\end{tabular}%
}
\end{table*}

\noindent
\textbf{Crafting the Benchmark.}
After comprehensively collecting technical reports from well-known blockchain security companies~\cite{blocksecteam2023}, we have identified six confirmed vulnerable contracts, as \textbf{P}. As all their source code files are available, we manually patch each of them to compose $\overline{\textbf{P}}$.
Moreover, we manually sample four benign contracts from widely-adopted DeFi, i.e., Aave~\cite{aave2023}, Compound~\cite{compound2023}, ParaSpace Lending~\cite{paraspace2023}, and a yield protocol~\cite{convexfinance2023}, to form the set \textbf{N}.
The reason of selecting these four contracts is twofold.
On the one hand, they all require the input of external contract address. Specifically, Aave and Compound primarily employ tokens as collateral to borrow another valuable token, while ParaSpace uses NFTs as collateral. The yield protocol uses the input external contract address to generate collateral yield. Because all these four contracts perform the necessary verification on the passed address, they meet both \textbf{P1} and \textbf{P2}.
On the other hand, after executing certain on-chain operations, the valuable tokens in their contracts are transferred to the caller, thus posing potential risks as \textbf{P3}.
Similarly, we deliberately remove their verification on addresses to make them vulnerable, denoting this set as $\overline{\textbf{N}}$.
Consequently, we obtained 20 ground truth cases.
Table~\ref{table:benchmark} illustrates the results, where the highlighted rows refer to the mis-detected results.

\noindent
\textbf{Average Time.}
It takes {\framework} around 7.34s on average, while Mythril lags considerably, taking about 28.3s on average. Ethainter and Jackal sit in between, with times ranging from 10.86s to 19.19s on average. {\framework} is approximately 3.86x, 1.48x, and 2.61x faster than Mythril, Ethainter, and Jackal, respectively.
When considering the timeout cases, both {\framework} and ETHBMC recorded zero, while Mythril, Ethainter and Jackal encountered timeout in 8, 4, and 6 instances, respectively.
Additionally, we can easily observe that ETHBMC performs well in terms of executing time. However, after a comprehensive code audit and analysis on output logs, we believe it is because its bounded model checking approach prioritizes efficiency over thoroughness. In other words, paths may be overlooked, which may compromise the accuracy in complex scenarios, like the address verification vulnerability focused in this work.

\noindent
\textbf{Precision \& Recall.}
Precision and recall are two critical metrics for evaluating an analyzer's effectiveness, where {\framework} outperforms other tools. Specifically, {\framework} achieves 100\% precision and 100\% recall on the benchmark. 
In the case of Mythril and ETHBMC, the main issue is false negatives. For the cases that can be completed within the time limit, Mythril and ETHBMC have a 50\% and 100\% false negative rate, respectively.
We speculate that the primary reason for ETHBMC's non-ideal results is twofold. 
On the one hand, ETHBMC's bounded model checking strategy inherently focuses on a specific range of states and paths within contracts, potentially missing the complexities involved in address verification due to its limited scope.
On the other hand, ETHBMC necessitates a pre-defined initial state for analysis. However, this state is very likely not optimal for detecting the address verification vulnerability, potentially affecting its performance.
Ethainter also has a worse performance compared to {\framework} in terms of recall, with a  false negative rate of around 12.5\%.
We think the most critical factor is the adoption of Gigahorse~\cite{grech2019gigahorse}, a toolchain for binary analysis. According to its implementation, one of its limitations is its inability to perfectly handle dynamic memory, affecting the performance of Ethainter in identifying functions that extensively use dynamic memory. Consequently, this limitation leads to the false negatives.

\noindent
\textbf{Root Causes.}
Considering the differences in metrics when conducting analysis on the benchmark among these five tools, we speculate that there are four reasons for their distinctions on the performance on the benchmark.
First, {\framework} fully leverages the characteristics summarized from \textbf{P1}. In the {\grapher}, it filters suspicious functions as candidates, which significantly reduces the number of possible states, a predicament affects these tools. 
Second, as detailed in \S\ref{sec:design:m2:control}, the path-searching strategy employed by the {\simulator} is specifically designed for the address verification vulnerability. This strategy prioritizes paths that may lead to vulnerabilities.
Thirdly, when handling the dynamic memory, the other four tools struggle to accurately analyze vulnerable functions that extensively employ complex dynamic memory allocation. In contrast, {\framework} leverages an EVM simulator, enabling it to precisely track address parameters without explicitly modeling dynamic memory behaviors, thereby enhancing its capability to identify such functions.
Last, {\framework} adopts a straightforward simulation approach, rather than static symbolic execution. This choice contributes to its efficiency. Previous studies, like KLEE~\cite{cadar2008klee}, suggest that backend SMT solvers can be significant drags on performance.

\subsubsection{Real-world Contracts Results}
\label{sec:evaluation:rq1:real-world}
To further illustrate the effectiveness of {\framework} on real-world contracts, we perform the analysis on all collected contracts, 5,158,101 ones in total.
Consequently, 812 of them are marked as vulnerable by {\framework}. 
To evaluate the effectiveness of {\framework}, we again use Mythril, Ethainter, Jackal, and ETHBMC as baselines.
However, because the unreadability of the bytecode, for a more effective comparison, we tried to obtain their source code from Etherscan.
Finally, we collected 369 pieces of source code.
The final scanning results for all these five tools on 369 open-source contracts are shown in Table~\ref{table:Open CA}.

\begin{table}[]
\centering
\caption{Performance comparison among {\framework}, Mythril, Ethainter, Jackal, and ETHBMC on real-world contracts.}
\label{table:Open CA}
\resizebox{\columnwidth}{!}{%
\begin{tabular}{cccccc}
\toprule
\textbf{Metrics} & \textbf{{\framework}} & \textbf{Mythril}* & \textbf{Ethainter}* & \textbf{Jackal}* & \textbf{ETHBMC}*\\ \midrule
Avg. Time(s)     & 6.34       & 33.69  & 8.74   &29.96   &5.43  \\
\rowcolor[HTML]{C0C0C0}
\# Timeout       & 0          & 42      & 6      &60     &164   \\
True Positives   & 348        & 16      & 147     &172   &2  \\
True Negative   & 0        & 2      & 4     &11    &21   \\
\rowcolor[HTML]{C0C0C0}
False Positives  & 21         & 8       & 17      &10    &0   \\
\rowcolor[HTML]{C0C0C0}
False Negatives  & 0          & 301     & 195     &116   & 182   \\ 
Precision  & 94.3\%          & 66.7\%     & 89.6\%     &94.5\%   & 100\%   \\ 
Recall  & 100\%          & 5.1\%    & 43.0\%     &59.7\%   & 1.1\%   \\ \bottomrule
\multicolumn{4}{l}{*The address vulnerability detector is implemented by ourselves.}                                         
\end{tabular}%
}
\end{table}

\noindent
\textbf{Average Time.}
As we can see, for all 369 cases, {\framework} achieves the second-best performance in terms of average analysis time. Moreover, there are no timeout cases within the 10-minute limit.
Ethainter is one place behind, with 8.74s and 6 timeouts.
Mythril's efficiency lags far behind, averaging 33.69s per case and suffering 42 timeouts exceeding the 10-minute threshold, the highest among compared tools.
Jackal averages around 29.96s with 60 timeouts within.
Finally, though ETHBMC has a decent 5.43s average time, it suffers 164 timeout cases, which significantly impacts its effectiveness.
By comparing Table~\ref{table:Open CA} and Table~\ref{table:benchmark}, we can observe that the performance among these tools is roughly consistent, except that ETHBMC has more timeout cases on the real world cases. We speculate that this is because it needs to try different initial states when the current one cannot explore paths to exploitations, leading to a huge efficiency issue.

\noindent
\textbf{Precision \& Recall.}
After manually rechecking all these contracts, the numbers of false positives and false negatives are also shown in Table~\ref{table:Open CA}.
We can easily observe that there are 21 false positives generated by {\framework}, leading to 94.3\% precision.
The main reason for that is there are \textit{unconventional verification methods on addresses}.
Except for the three mechanisms we summarized in \S\ref{sec:background:verification}, some of them delegate address verification to other contracts, which is not a widely adopted verification method. Moreover, some contracts perform verification via digital signatures~\cite{buterin2016ethereum} or Merkle proofs~\cite{liu2021merkle}.
Currently, due to efficiency issues, {\framework} does not integrate such patterns. Moreover, as inter-contract analysis is always a huge obstacle for smart contract analysis~\cite{ma2021pluto}, it is a compromise must be made.
In contrast, all the other four baselines suffer from severe false negative issues.
Although Mythril employs a similar path filtering approach to {\framework}, its recall is only 5.1\%, because the symbolic execution cannot effectively find feasible paths to exploit vulnerable contracts.
Ethainter and Jackal, both of which use the GigaHorse framework, achieve recall rates of only 43.0\% and 59.7\%, respectively.
As we mentioned in \S\ref{sec:evaluation:rq1:benchmark}, Gigahorse struggles to accurately construct complete CFGs when handling some contracts, which stems from its less optimized handling on the dynamic memory.
The recall of ETHBMC is only 1.1\%, whose reason is mainly due to its adopted initial states as we stated above.
We have conducted a case study to illustrate how a case is mislabeled as false negative by these four tools, please refer to our open-source repo at \href{https://github.com/avverifier/avverifier}{link}.

\begin{tcolorbox}[title= \textbf{RQ-1} Answer, left=2pt, right=2pt, top=2pt,bottom=2pt]
Compared to Mythril, Ethainter, Jackal, and ETHBMC, {\framework} can improve the efficiency 2 to 5 times. {\framework} can achieve at least 94\% precision and 100\% recall on the well-constructed benchmark and real-world contracts, while the others suffer a severe false negative issue due to their design and implementation.
\end{tcolorbox}

\subsection{RQ2: Characteristics of Real-world Vulnerable Contracts}
\label{sec:evaluation:rq2}
To characterize the ecosystem of real-world vulnerable contracts, we firstly illustrate the overall results on all vulnerable ones among over 5M collected ones (see \S\ref{sec:evaluation:rq2:overall}).
Then, we focus on the behavioral characteristics, including \textit{financial related} and \textit{activity related}, of these vulnerable contracts (see \S\ref{sec:evaluation:rq2:activity} and \S\ref{sec:evaluation:rq2:financial}).

\subsubsection{Overall Results}
\label{sec:evaluation:rq2:overall}
As we mentioned in \S\ref{sec:evaluation:rq1:real-world}, we have applied {\framework} on all 5.2M contracts.
In total, we have identified 812 vulnerable ones.
Among them, we found 443 of them are close-sourced. According to the MD5 results, we have successfully identified 131 unique close-sourced contracts. To recheck the identified results, we decompiled the unique close-sourced ones by contract library tools~\cite{dedaub_decompile}, a well-known decompiler, and asked two Ph.D. students who major in this area to recheck the results. The manual recheck has not revealed any false positives.
Due to the unreadability of the close-sourced bytecode, even for the decompiled ones, we can only confirm that 17 of close-sourced contracts are related to Ethereum tokens according to the function signature.

As for the remaining 348 true positive cases mentioned in Table~\ref{table:Open CA}, we obtained their source code from Etherscan.
Similarly, by deduplicating the source code, we finally obtained 299 unique ones. After a manual recheck, the classification results are shown in the Table~\ref{table:Classification}.
We can observe that nearly half of them are related to ERC-20, indicating potential financial impacts of this vulnerability. Moreover, around one-third are in the DApp category, like lending market or swap router, which may also bring in impacts to the Ethereum ecosystem.

We believe that these open-source contracts can gain more user trust and are more representative.
In the following \S\ref{sec:evaluation:rq2:activity} and \S\ref{sec:evaluation:rq2:financial}, we characterize these 348 open source smart contracts, as a lower-bound of the overall landscape.

\begin{table}[]
\centering
\caption{For 348 identified true positive open-source cases, the classification results according to their functionalities.}
\label{table:Classification}
\resizebox{0.8\columnwidth}{!}{%
\begin{tabular}{ccccc}
\toprule
\textbf{Contract Type}           & \textbf{ERC-20} & \textbf{ERC-721} & \textbf{DApp} \\ \midrule
Open-source Contract             & 177            & 61              & 110           \\
Unique Open-source Contract      & 153            & 45              & 101           \\ \bottomrule
\end{tabular}%
}
\end{table}

\begin{figure}
\centering
  \includegraphics[width=1\linewidth]{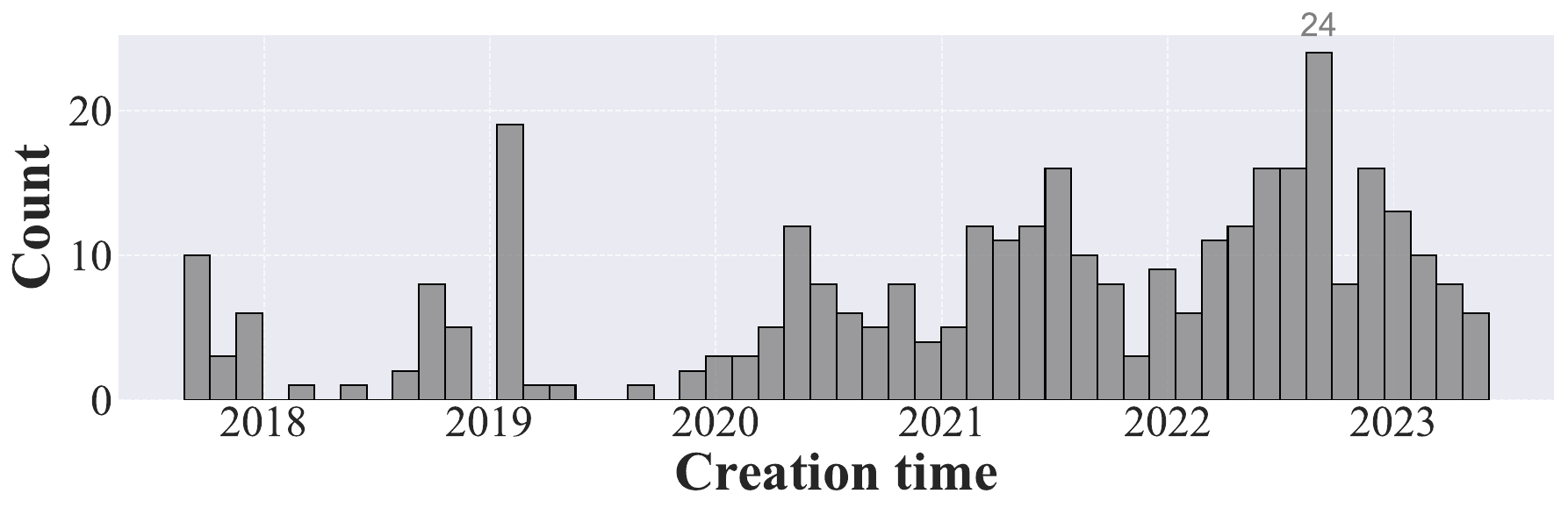}
         \caption{Distribution of vulnerable ones by creation time.}
         \label{fig:Vulnerable contract timestamp histogram}
\end{figure}%

\begin{figure}[t]
  \centering
  \includegraphics[width=1\linewidth]{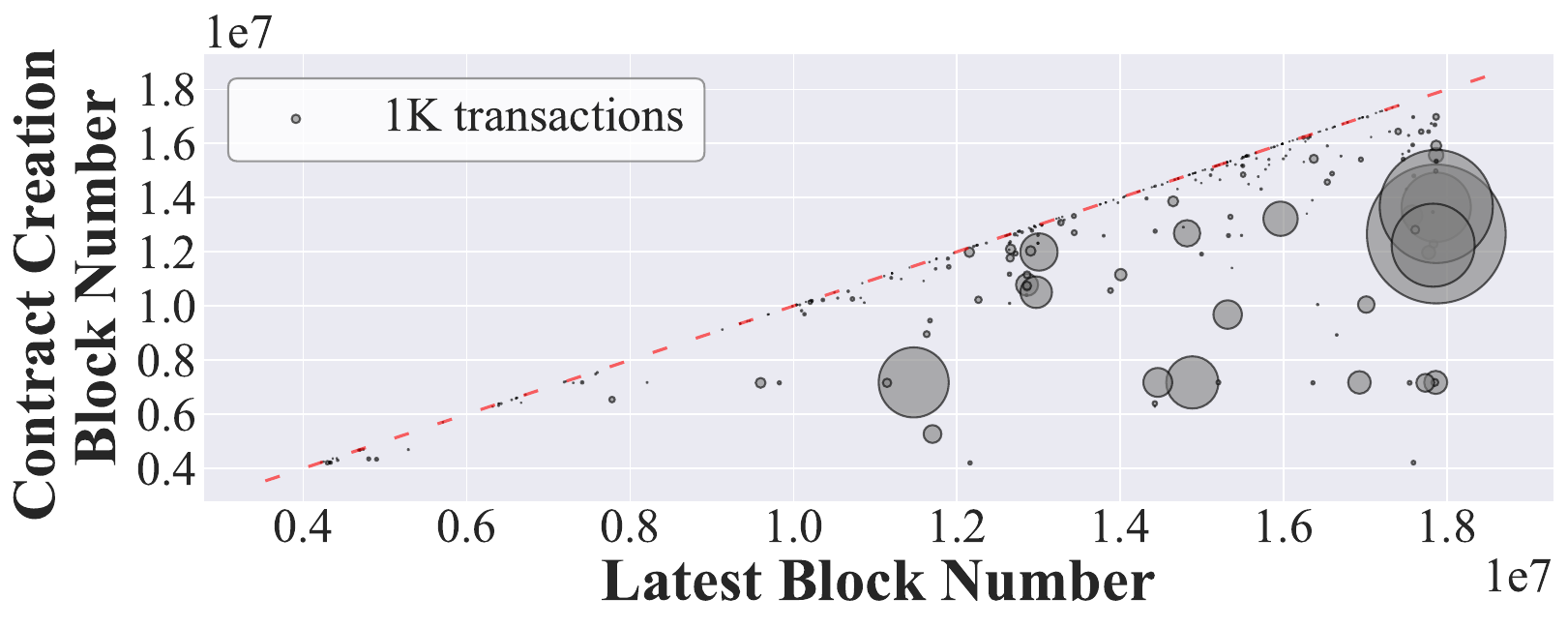}
         \caption{Relations between creation time and lifespan for vulnerable contracts.}
         \label{fig:Bubble Chart}
\end{figure}

\subsubsection{Activity Related Metrics}
\label{sec:evaluation:rq2:activity}
To depict the activity of them, we first illustrate the distribution of their deployment time, as shown in Fig.~\ref{fig:Vulnerable contract timestamp histogram}.
As we can see, the figure illustrates a general upward tendency, which aligns with our intuition as Ethereum has been growing and the propagation of a vulnerability is a temporal phenomenon.
There are two noticeable troughs in late-2019 and late-2021. Delving deeper, we discerned that the first trough was primarily influenced by stringent cryptocurrency policies, leading to a liquidity crunch. Meanwhile, the second one can be attributed to the aftermath of the \textit{Luna Event}~\cite{liu2023anatomy}, which precipitated a significant liquidation of assets, further causing a depletion in liquidity.
We can also observe a peak located in the January 2019.
This is because the birth of Uniswap~\cite{uniswap2023} in November 2018, which played a pivotal role in the DeFi prosperity, leading to a substantial increase in the number of contract creations as well as the vulnerable ones.

We also observe the lifespan of these 348 vulnerable smart contracts in Fig.~\ref{fig:Bubble Chart}, where the y-axis and x-axis represent the block height when a contract was created and the time when the last transaction occurred respectively. The size of the bubble is proportionally to its historical transaction count.
As we can see, there are lots of tiny bubbles locate along or near the diagonal red line. This implies that a significant number of contracts had a very short lifespan. Such a trend suggests a plethora of transient contracts, potentially due to testing exercises, spamming campaigns, or temporary endeavors within the Ethereum ecosystem.
Interestingly, while larger bubbles scattered across the figure signify contracts with considerable transactional activity, they may still be susceptible to the address verification vulnerability. 
We speculate their persistent activity suggests that they may correspond to low balance, which likely diminishes the motivation for attackers to exploit their vulnerabilities.
Additionally, we observe some large bubbles distribute near the red line. This indicates a burst of transactions in their earlier phases but have since transitioned into a dormant state. Such an intriguing contrast between their past vibrancy and current inactivity prompts us to further investigate these contracts.
Therefore, we have filtered out the transactions of the top 50 contracts in terms of the number of involved transactions, i.e., the size of bubbles.
Through a detailed and comprehensive transactional analysis, we found that 3 of them have been exploited already, while 47 of them are at risk.
We speculate the reason is that the balance is small and attackers have not noticed yet or are waiting for the opportunity to make a large profit.

\begin{figure}
\centering
  \includegraphics[width=1\linewidth]{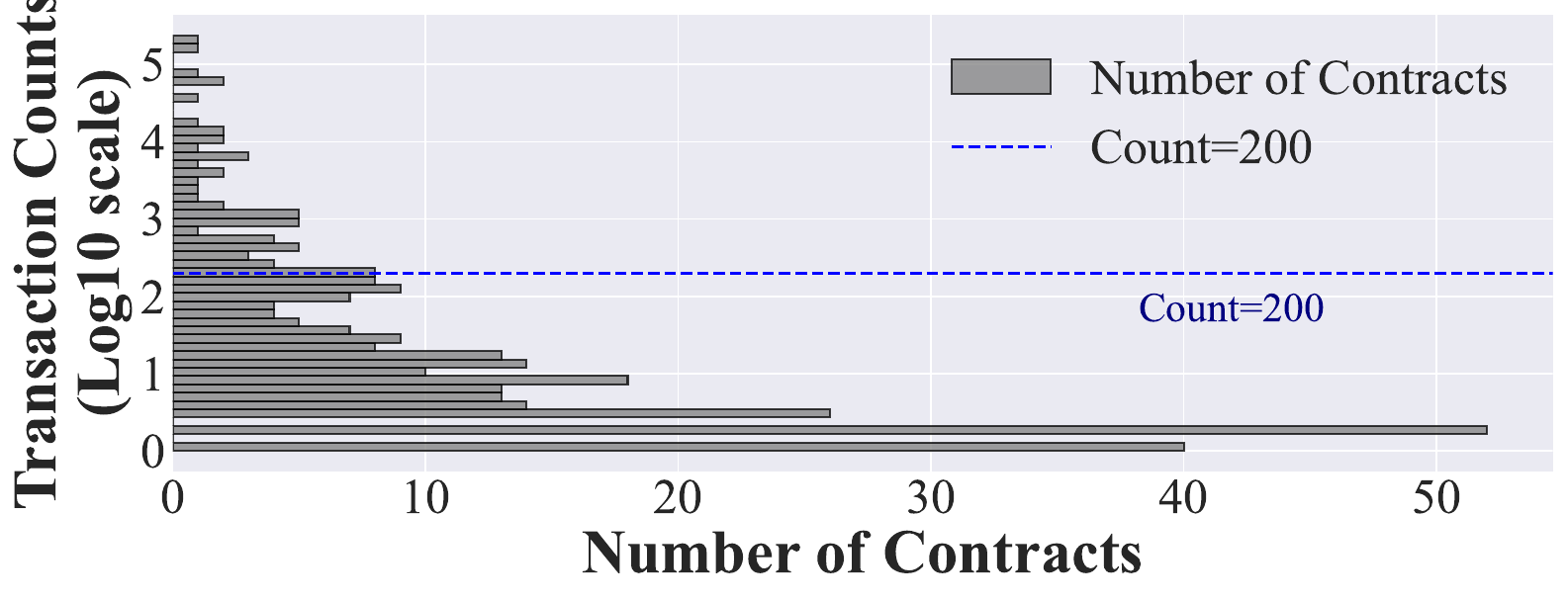}
  \caption{Distribution of the number of transactions involved in vulnerable contracts.}
  \label{fig:Histogram tx Counts}
\end{figure}%

\subsubsection{Financial Related Metrics}
\label{sec:evaluation:rq2:financial}
We further evaluate the financial impact of these vulnerable smart contracts.
Intuitively, by measuring how many transactions are involved in a contract can reflect its financial impact to some extent.
Fig.~\ref{fig:Histogram tx Counts} illustrates the distribution of the number of transactions involved in them.
We can easily observe that the distribution follows the Pareto principle~\cite{dunford2014pareto}, i.e., there exists a long tail in the distribution.
More than 55.46\% cases are involved within 5 transactions, while the case named AnySwap~\cite{anyswap2023} is involved in around 237,000 transactions.
The results follow the Oliva et al.'s~\cite{oliva2020exploratory}, i.e., most contracts in Ethereum are inactive, a small portion of contracts greatly contribute to the prosperity of the Ethereum ecosystem.

In addition, we desire to evaluate how many assets are directly involved in them.
For DeFi projects, TVL (Total Value Locked) is one of the most representative metrics.
Thus, for each case, we retrieve its historical TVL from a third-party API, DeFiLlama~\cite{defillama2023ethereum}, a DeFi data browser known for providing metrics like TVL and market cap.
For each project, we retrieve their peak TVL as it corresponds to the timestamp when attackers can obtain the most profit.
In total, around \$11.2 billion was considered to be directly locked into these vulnerable contracts.
The top-3 Ethereum projects in terms of compromised assets include Visor\footnote{0xC9f27A50f82571C1C8423A42970613b8dBDA14ef}~\cite{beosin2023visor}, TempleDao\footnote{0xd2869042E12a3506100af1D192b5b04D65137941}~\cite{templedao}, and AnySwap\footnote{0x6b7a87899490EcE95443e979cA9485CBE7E71522}~\cite{anyswap}.
Interestingly, except for Anyswap, other two projects immediately stop providing services after being attacked.
This is because hackers exploit a vulnerability in Anyswap to indirectly steal tokens from users. Conversely, in the other two projects, the funds within the vulnerable contracts were directly accessed and stolen, leading to an immediate cessation of services.
Moreover, the time window between the creation and the corresponding exploitation suggests the opportunity for vulnerability detection and remediation. While Visor saw exploitation within 3 months, Anyswap remained uncompromised for 210 days. Such a variation might result from factors like the implementation complexity, public visibility, or the inherent vulnerability's nature.
However, it proves that there is often a time window to detect and patch vulnerabilities before an attack happens.

\begin{tcolorbox}[title= \textbf{RQ-2} Answer, left=2pt, right=2pt, top=2pt,bottom=2pt]
Around 68.4\% vulnerable contracts are ERC-20 or ERC-721 tokens. We further reveal that attackers tend to launch attacks dozens of days after the deployment for greater benefits, suggesting there exists a time window for vulnerability detection and remediation.
\end{tcolorbox}

\subsection{RQ3: Real-time Detection}
\label{sec:evaluation:rq3}
We aim to deploy {\framework} as a real-time detector.
Thus, we measured several performance metrics (see \S\ref{sec:evaluation:rq3:efficiency}).
Additionally, we give a case study to illustrate how a vulnerability can be detected by {\framework} before an attack (see \S\ref{sec:evaluation:rq3:case}).

\subsubsection{Quantitative Analysis}
\label{sec:evaluation:rq3:efficiency}
We measure two real-world performance metrics.
First, we compare the rate of contract creation along block generation to the performance of {\framework}, which can shed light on the responsiveness and real-time applicability of {\framework}.
Second, we illustrate the correlation between bytecode length and the consumed time taken for analysis. This metric indicates the scalability of {\framework} with the increasing complexity and size of contracts.

\noindent
\textbf{Rate on Contract Creation vs. Detection.}
We directed our attention to the data from Nov. 2022 to Jan. 2023, a period of time when contracts are heavily deployed (illustrated in Fig.~\ref{fig:Vulnerable contract timestamp histogram}).
According to our statistics, these three months cover blocks with the height from 15,870,000 to 16,518,000, accounting for 289,238 deployed contracts. In other words, 0.45 contract is deployed on average within each block.
As for BSC, according to a widely known BSC browser, BscScan~\cite{bscscan2023}, we can calculate that a block is generated every 3s, and there are 2.1 contracts deployed in each BSC block on average.
According to the results in RQ1, each Ethereum contract takes around 6.42s. Therefore, considering the number of contract deployed in each block and the speed of block generation in Ethereum, a single-core processor can be used to deploy {\framework} as a real-time detector.
As for BSC, each block spends around $6.42s\ \times 2.1 = 13.48s$, greater than the time taken by the block generation. However, the methodologies adopted by {\framework} can be paralleled easily, like analyzing multiple suspicious functions simultaneously. Therefore, a multi-core machine is sufficient.

\begin{figure}
  \centering
  \includegraphics[width=1\linewidth]{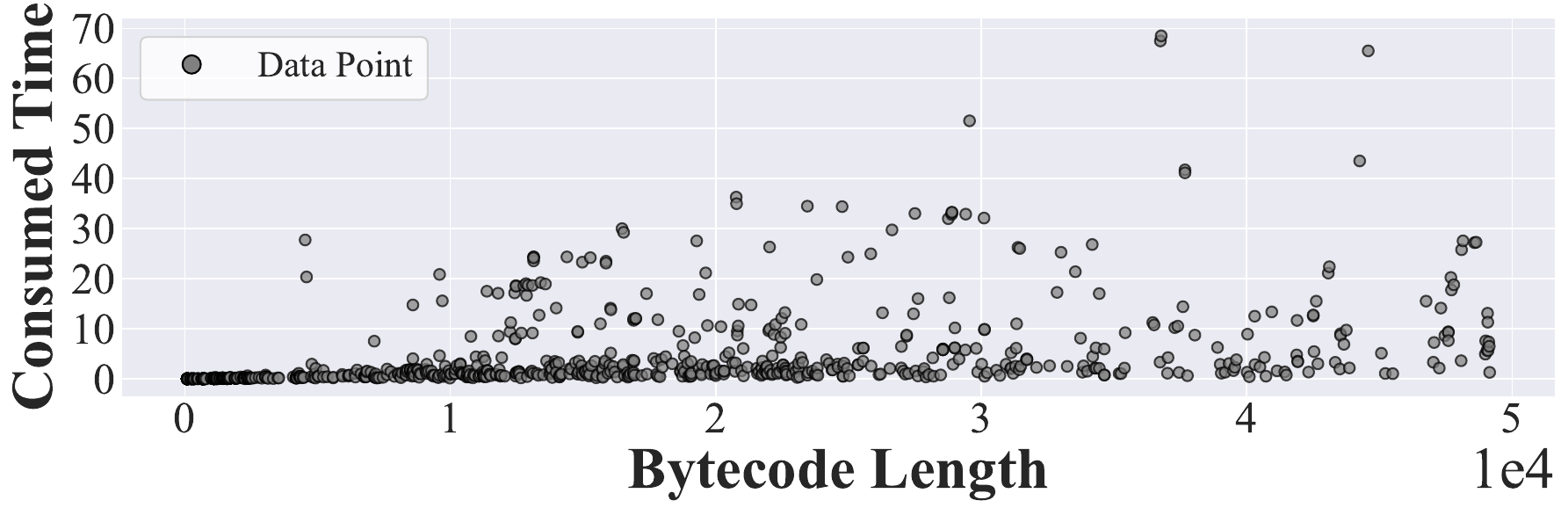}
    \caption{The relationship between the bytecode length and the time consumed on each case.}
    \label{fig:run_time}
\end{figure}

\noindent
\textbf{Scalability.}
To evaluate the scalability of {\framework}, we randomly sample 1,000 contracts from the ones deployed within the recent year.
Figure~\ref{fig:run_time} presents the relation between the bytecode length and the consumed time.
Clearly, there does not exist a linear correlation or even an exponential one between these two metrics. We can also observe that most cases can be finished within 20s.
Such a high detection efficiency can be attributed to two points.
On the one hand, the detection logic is very efficient. For example, the {\detector} can effectively screen suspicious functions, and can stop the analysis in time when a vulnerability is encountered.
On the other hand, the detection method has few performance bottlenecks. Unlike static symbolic execution techniques, the {\simulator} can quickly and accurately traverse paths that could lead to vulnerabilities.
Therefore, the spent time of {\framework} on each case is not directly proportional to bytecode length, illustrating its scalability.

\begin{listing}[t]
\begin{minted}{java}
function 0xbc423deb(uint256 varg0, uint256 varg1, uint256 varg2) public payable { 
    require((address(varg2)).code.size);
    v0 = address(varg2).slip(_ilk, msg.sender, 0 - varg1).gas(msg.gas);
    require(v0); // checks call status, propagates error data on error
    v1, v2 = _gem.transfer(address(varg0), varg1).gas(msg.gas);
    require(v1); // checks call status, propagates error data on error
    emit Exit(address(varg0), varg1);
}
\end{minted}
        \vspace{-0.1in}
    \caption{A case that is detected by our real-time detector.}
    \vspace{-0.1in}
    \label{lst:para}
\end{listing}

\subsubsection{Case Study: A Real-world Early-warning Case}
\label{sec:evaluation:rq3:case}
We illustrate a real-world case that is marked as vulnerable when {\framework} is deployed as a real-time detector on BSC.
As there is no source code for the case, Listing~\ref{lst:para} illustrates its decompiled version.
Moreover, due to the non-disclosure principle, we make a slight change syntactically without modifying its original semantics.

As we can see, L2 indicates that the \texttt{varg2} is an address passed from the external environment (satisfying \textbf{P1}).
At L3, the function invokes \texttt{slip}, which takes the \texttt{varg2} as the target address (satisfying \textbf{P2}).
Then, at L4, a \texttt{require} checks the returned value (satisfying \textbf{P3}).
Finally, at L5, the contract invokes \texttt{transfer} to transfer tokens to the address referred by \texttt{varg0} .
Therefore, attackers can deploy a contract to bypass the verification on \texttt{v0} to drain this contract out.

Notably, we detected this vulnerability on May 18th, 2023, at 6:10 UTC. An attack transaction was initiated 1.5 hours after, at 7:41 UTC.
Such a time gap highlights that capability of {\framework}.
However, the absence of an automated exploit response mechanism within {\framework} prevented timely intervention, leading to a user loss of \$ 30K USD.
This incident underscores the importance of designing an automated response tool to mitigate potential financial damages resulted from the address verification vulnerability.

\begin{tcolorbox}[title= \textbf{RQ-3} Answer, left=2pt, right=2pt, top=2pt,bottom=2pt]
Though a quantitative analysis, it illustrates the usability and scalability of {\framework} as being a real-time detector. A real-world case, worth approximately \$30,000, proves its ability to raise early-stage warnings.
\end{tcolorbox}

%% file: sec-conclusion.tex
\section{Discussion}
\label{sec:discussion}

\subsection{Threats to Validity}

\noindent
\textbf{Scalability of {\framework}.}
{\framework} is specifically designed for the address verification vulnerability. However, due to the efficient taint analysis we implemented in {\framework}, 
it can be easily extended to detect other kinds of vulnerabilities, 
like unchecked external calls and inadequate access controls.
Moreover, {\framework} can also be adopted on other blockchain platforms, i.e., EVM-compatible chains, like BSC, TRON, and AVAX.
Consequently, {\framework} is not a tool limited to the address verification vulnerability on Ethereum.

\noindent
\textbf{Candidate Contracts.}
When answering RQ1 and RQ2, we only choose around 5 million contracts as candidates.
According to the statistics, there still exist millions of contracts. However, we think those already deployed ones without transactions are the worthless targets for attackers.
However, in answering RQ3, i.e., deploying {\framework} as a real-time detector, we take all newly deployed contracts as candidates.

\subsection{Limitations}
Although the {\framework} is effective in identifying the address verification vulnerability, it still faces the following three limitations:

\noindent
\textbf{Dynamic Parameters Verification Mechanism.} Some smart contracts employ dynamic parameters verification mechanisms, like digital signatures, which require loading data into memory. To this end, the size of \texttt{CALLDATA} is unknown beforehand, posing challenges to {\framework} as the variable length of \texttt{CALLDATA} leads to unpredictable offsets, making it difficult to track variable arguments in dynamic memory.
Currently, for such dynamic parameters, {\framework} concretizes their length to obtain the final results. Due to its possibility of introducing false negatives, we take this as one of our future research directions.

\noindent
\textbf{Scalability.} {\framework} is a tool specifically designed to identify the address verification vulnerability. As such, currently, it cannot be used for identifying other smart contract vulnerabilities directly.
However, because we have implemented a comprehensive set of taint propagation rules, {\framework} can be easily extended to other smart contract security problems.

\noindent
\textbf{Auto-exploitation.} While {\framework} can efficiently and accurately identify the address verification vulnerability and the entry point, it cannot automatically generate exploitation yet. Currently, all identified contracts still require a comprehensive manual confirmation to determine if they are truly vulnerable and exploitable.
However, we can take advantage of other techniques, like fuzzing and symbolic execution, to help {\framework} generate exploitations and \texttt{CALLDATA} based on the generated results. We believe it is a meaningful research area.

\subsection{Ethical Consideration}
Due to the anonymity of blockchain, it is only possible to
contact the developers of open-sourced projects.
Thus, we have tried our best to contact developers of vulnerable contracts through all possible and usable media, like mail, official website, and twitter, with more
than 500 transactions in history. Unfortunately, after we warned 48 developers once after the identification and manual recheck, we did not receive any responses from them within 2 weeks.
Where we deploy a real-time detector on Ethereum and BSC, we observed that {\framework} has raised alarms several times, like the one shown in \S\ref{sec:evaluation:rq3:case}. However, it is impossible to reach its developers privately. Moreover, we cannot initiate a transaction to notify the developer, because it will also notify malicious users that there is a possible victim.
Thus, we urge collaboration with blockchain security companies for the secure and efficient disclosure of vulnerabilities. 

\vspace{-0.1in}
\section{Related Work}
\label{sec:related work}

\noindent
\textbf{Smart Contract Vulnerability Detection.}
Vulnerability detection in smart contracts employs varied methodologies dependent on input types and detection principles. Analyzing contracts can either occur at the high-level source code \cite{gao2019easyflow, ma2019evm} or through the bytecode interfacing with the EVM \cite{albert2018ethir, grishchenko2018ethertrust}.
Further categorizing based on analytical techniques, static analysis delves into code structure and inherent semantics to identify vulnerabilities, often applied to source code evaluations. Dynamic analyses, more prevalent in bytecode assessments, utilize strategies such as fuzz testing to spot anomalies by bombarding contracts with randomized inputs \cite{ashraf2020gasfuzzer, jiang2018contractfuzzer}. 
In addition to these, hybrid analysis methods are also gaining traction, combining static and dynamic analysis techniques for a more thorough examination of smart contracts. 
Through this combined approach, hybrid analysis helps in identifying a wider range of vulnerabilities \cite{li2021hybrid, ma2023pied}. 
Furthermore, trace-based evaluations provide a unique perspective by scrutinizing historical transaction patterns, unearthing vulnerabilities from real-world usage patterns \cite{zhang2020txspector, chen2019dataether}.

\noindent
\textbf{Taint Analysis in Smart Contracts.}
Pioneering efforts in taint analysis have led to the development of tools tailored specifically for Ethereum contracts, adept at detecting common vulnerabilities such as unchecked send and reentrancy \cite{alkhalifah2021mechanism, rodler2018sereum}. Other research efforts have heightened the granularity of taint analysis to uncover intricate data leaks and permission oversights \cite{tikhomirov2018smartcheck, torres2018osiris}. Furthermore, several advanced methods have integrated both static and dynamic analysis techniques, blending the advantages of both to provide a more comprehensive security assessment \cite{gao2019easyflow, kushwaha2022ethereum}. These collective advancements underscore the pivotal role of taint analysis in shaping a robust ecosystem for smart contracts.

\noindent
\textbf{Permission Checks and Access Control.}
Permission checks underpin smart contract security, preventing unauthorized actions which can lead to financial or data losses \cite{chen2021smart, liu2022finding}. Many tools and frameworks now aid developers in verifying permissions \cite{shahab2020reducing, kamboj2021user}. However, some advanced attacks can bypass traditional checks, underscoring the need for context-aware analysis \cite{zhang2018smart, cruz2018rbac}. Rigorous permission validation remains crucial for a secure smart contract environment.

\section{Conclusion}
\label{sec:conclusion}
In this work, we present {\framework}, a taint analyzer based on static EVM opcode simulation, which is designed for identifying the address verification vulnerability hidden in Ethereum smart contracts.
With the help of the heuristic-based path selection method and taint propagation rules in {\simulator}, as well as the three-phase formal detection rules implemented in {\detector}, {\framework} significantly outperforms Mythril in both terms of efficiency and effectiveness.
According to a comprehensive evaluation on over 5 million contracts, as well as the behaviour characteristics they illustrate, it proves the necessity of implementing {\framework}.
Additionally, {\framework} is proven efficient and effective enough to be a real-time detector on EVM-like blockchain platforms to raise early warnings once contracts are deployed.

\section*{Availability}
We have released {\framework} and the benchmark at \href{https://github.com/avverifier/avverifier}{link}.